\documentclass[aps,twocolumn,showpacs,prl,10pt]{revtex4-1}

\usepackage{epsfig,amsmath,color,colordvi,verbatim}
\usepackage{amssymb,amsfonts,amsmath}

\begin{document}


\def\be{\begin{equation}}
\def\ee{\end{equation}}
\def\bee{\begin{eqnarray}}
\def\eee{\end{eqnarray}}
\def\sech{\mbox{sech}}
\def\e{{\rm e}}
\def\d{{\rm d}}
\def\L{{\cal L}}
\def\U{{\cal U}}
\def\M{{\cal M}}
\def\T{{\cal T}}
\def\V{{\cal V}}
\def\R{{\cal R}}
\def\kb{k_{\rm B}}
\def\tw{t_{\rm w}}
\def\ts{t_{\rm s}}
\def\Tc{T_{\rm c}}
\def\gs{\gamma_{\rm s}}
\def\tm{tunneling model }
\def\TM{tunneling model }
\def\tilde{\widetilde}
\def\Deltac{\Delta_{0\rm c}}
\def\Deltamin{\Delta_{0\rm min}}
\def\Emin{E_{\rm min}}
\def\tauc{\tau_{\rm c}}
\def\tauac{\tau_{\rm AC}}
\def\tauw{\tau_{\rm w}}
\def\taumin{\tau_{\rm min}}
\def\taumax{\tau_{\rm max}}
\def\de{\delta\varepsilon / \varepsilon}
\def\pF{{\bf pF}}
\def\pFAC{{\bf pF}_{\rm AC}}
\def\halb{\mbox{$\frac{1}{2}$}}
\def\dreihalb{\mbox{$\frac{3}{2}$}}
\def\viertel{\mbox{$\frac{1}{4}$}}
\def\achtel{\mbox{$\frac{1}{8}$}}
\def\with{\quad\mbox{with}\quad}
\def\und{\quad\mbox{and}\quad}
\def\za{\sigma_z^{(1)}}
\def\zb{\sigma_z^{(2)}}
\def\ya{\sigma_y^{(1)}}
\def\yb{\sigma_y^{(2)}}
\def\xa{\sigma_x^{(1)}}
\def\xb{\sigma_x^{(2)}}
\def\spur#1{\mbox{Tr}\left\{ #1\right\}}
\def\erwart#1{\left\langle #1 \right\rangle}
\def\tr{{\rm tr}}
\newcommand{\bbbone}{{\mathchoice {\rm 1\mskip -4mu l}{\rm 1\mskip 
-4mu l}{\rm 1\mskip -4.5mu l}{\rm 1\mskip -5mu l}}}

\newcommand{\bj}{{\bf j}}
\newcommand{\bx}{{\bf x}}

\newcommand{\pn}[1]{{\color{red}{#1}}}

\title{Vibronic speed-up of the excitation energy transfer in the
Fenna-Matthews-Olson complex}

\author{P. Nalbach, C. A. Mujica-Martinez, and M. Thorwart}
\affiliation{I.\ Institut f\"ur Theoretische Physik,  Universit\"at
Hamburg, Jungiusstra{\ss}e 9, 20355 Hamburg, Germany\\
The Hamburg Centre for Ultrafast Imaging, Luruper Chaussee 149, 22761 Hamburg, Germany} 
\date{\today}


\begin{abstract} 

We show that the efficient excitation energy transfer in the
Fenna-Matthews-Olson molecular aggregate under realistic physiological
conditions is fueled by underdamped vibrations of the embedding proteins. For
this, we present numerically exact results for the quantum dynamics of the
excitons in the presence of nonadiabatic vibrational states in the
Fenna-Matthews-Olson aggregate employing a environmental fluctuation spectral
function derived from experiments. Assuming the prominent 180 cm$^{-1}$
vibrational mode to be underdamped, we observe, on the one hand, besides
vibrational coherent oscillations between different excitation levels of the
vibration also prolonged electronic coherent oscillations between the initially
excited site and its neighbours. On the other hand, however, the underdamped
vibrations provide additional channels for the excitation energy transfer and by
this increase the transfer speed by up to $30 \%$.

\end{abstract}

\maketitle



Recent experiments on the ultrafast exciton dynamics in photoactive biomolecular complexes 
have sparked renewed interest in the longstanding question whether nontrivial quantum coherence 
effects exist in natural biological systems under physiological conditions, and, if so, whether 
they have any functional significance. Photosynthesis \cite{PhotosyntheticExcitons-book-2000} 
starts with the harvest of a photon by a pigment and the formation of a tightly bound 
electron-hole pair. After the exciton has been formed in the antenna complexes, its 
energy is transferred to the reaction center (RC), where further charge separation is 
initiated. The transfer dynamics of the excitations occurs non-radiatively by a Coulomb 
dipolar coupling and has traditionally been regarded as an incoherent hopping between
molecular sites \cite{May-Kuehn-2011}. 

Recently, Engel et al.\ \cite{Engel-Nature-2007,Engel-PNAS-2010} have interpreted long-lasting 
beating signals in time-resolved optical two-dimensional spectra of the Fenna-Matthews-Olson (FMO) 
complex \cite{FMO-structure-2_8A-JMB-1979,Brixner-Nature-2005, opticalpropertiesFMO-review-2010} as 
evidence for quantum coherent energy transfer via delocalized exciton states. The FMO complex is a 
trimer with seven bacteriochlorophyll (BChl) molecular sites (plus an additional weakly coupled eighth 
BChl) in each monomer. Quantum coherence times of almost a picosecond at $77$ K \cite{Engel-Nature-2007,Engel-PNAS-2010} 
and about $300$ fs at physiological temperatures \cite{Engel-PNAS-2010} have been reported.  

These reports have boosted on-going research to answer the question how quantum coherence can prevail 
over such long times in a strongly fluctuating physiological environment formed by the strongly vibrating 
protein host and the surrounding polarization fluctuations of the ambient water as a strongly polar 
solvent. Theoretical modeling of the real-time quantum transfer dynamics is difficult due to the 
non-standard spectral distributions of the fluctuations. In this case, the validity of standard 
Redfield-type approaches based on a weak-coupling and a Markovian approximation is not clear a 
priori \cite{Ishizaki-Fleming-JCP-2009,Nalbach-JCP-2010,Nalbach-NJP-2011}. Simplistic modelling 
shows that spatial and temporal correlations can in principle allow for prolonged quantum 
coherence \cite{Thorwart-CPL-2009,Nalbach-NJP-2010}. A more accurate treatment of the FMO model 
with seven localized sites and a physical environmental spectrum \cite{Wendling-JPCB-2000,Adolphs-Renger-BJ-2006} 
including a strongly localized Huang-Rhys vibrational mode \cite{May-Kuehn-2011} at thermal equilibrium has 
been realized. However, the resulting coherence times are considerably shorter than experimentally 
observed \cite{Nalbach-PRE-2011}. Very strong underdamped high-frequency vibrations could also be ruled 
out as a possible origin of the experimentally observed long-lived coherent beatings \cite{Nalbach-JPB-2012}. 

Recently, it was proposed that underdamped low frequency vibrations, which are strongly coupled to the 
electronic transitions with large Huang-Rhys factors, and which are almost resonant to the excitonic 
transitions, are responsible for the observed long-lived coherence \cite{Kauffmann-JPCB-2012,Kauffmann-JPCL-2012,Jonas-PNAS-2013,Plenio-NatPhys-2013}. 
The prominent vibrational mode with a wave number of $180$ cm$^{-1}$ was explicitely included in the 
system's nonadiabatic quantum dynamics and thus discussed on an equal footing as the excitonic states. 
By this, Christensson et al.\ \cite{Kauffmann-JPCB-2012} found coherence times in line with experimental 
findings employing a Redfield approach to treat the environmental fluctuations. Chin et al.\ \cite{Plenio-NatPhys-2013} 
observed the same results employing a numerically exact treatment of a partial model which is restricted to an effective 
FMO model with two of the seven pigments. So far, the full seven site FMO model with the realistic spectrum of its 
environmental fluctuations is unexplored by numerically exact means. Likewise, the fundamental question how long 
the quantum coherent beatings survive and how they profit from an interaction with the underdamped low frequency 
vibration in a realistic setting is open. We aim here at clarifying whether this scenario of excitonic coherence 
being fueled by vibrational coherence could explain the experiments on an accurate and quantitative level. At the 
same time possible functional relevance of the underdamped low frequency vibration is not studied so far. We 
investigate here the influence of the underdamped low frequency vibration on the speed and thus efficiency of 
energy transfer through the FMO complex.

In this work, we present numerically exact results for the full FMO model with seven sites, the spectral density 
of Refs.\ \cite{Wendling-JPCB-2000,Adolphs-Renger-BJ-2006} as extracted from optical spectra \cite{Wendling-JPCB-2000} 
and by treating the underdamped vibrational mode with wave number $180$cm$^{-1}$ explicitly and at an equal footing 
as the excitonic quantum dynamics. Upon adopting the iterative real-time  quasiadiabatic propagator path-integral 
(QUAPI) \cite{Makri-JCP-1995,Thorwart-CP-1998,LZ-Nalbach-Thorwart-PRL-2009} scheme we find  that the excitonic coherence 
times are increased by the coherent vibrational modes. We observe not only vibrational coherent oscillations between 
different excited vibrational states, but also prolonged electronic coherent oscillations between the initially excited 
site and its neighbours. Most importantly, we show quantitatively that the underdamped vibrational mode provides additional 
transfer channels for the excitation energy to be funnelled more efficiently through the Fenna-Matthews-Olson molecular 
aggregate under realistic physiological conditions. We find a substantial speed-up of the transfer times of up to $30 \%$, 
as compared to the static complex. Although both coherent oscillations and vibronic speed-up originate from the underdamped 
vibrational mode, the life time of the oscillations is irrelevant for the speed-up resulting in a robust mechanism.


\section*{Results}
\subsection*{Model of the FMO complex}

The FMO monomer contains eight bacteriochlorophyll {\it a} (BChl{\it a}) 
molecular sites \cite{opticalpropertiesFMO-review-2010}, of which the recently
discovered eighth pigment \cite{FMO-structure-1_3A-PR-2009,Renger-JPCL-2011} 
is only weakly coupled to the other BChls and thus will be omitted in the
present investigation. The Hamiltonian restricted to the single excitation subspace for the remaining seven sites is
$H_{FMO} = \sum_{j=1}^7 \epsilon_j |j\rangle\langle j| + 
           \sum_{j \neq k} J_{jk} \left( |j\rangle\langle k| + |k\rangle\langle j| \right)$,
where the basis states $|j\rangle$ indicate that the $j$-th site is in its 
excited state and all other sites are in their ground states. $\epsilon_j$ 
denotes the energy of the $j$-th site and $J_{jk}$ denotes the electronic
coupling between sites $j$ and $k$. We use the numerically determined site 
energies and dipolar couplings calculated by Adolphs and Renger 
\cite{Adolphs-Renger-BJ-2006} (the explicit form is given in the Supporting 
Information). Here, the BChl 3 is the site with the lowest energy which is connected to the RC and forms the exit site \cite{opticalpropertiesFMO-review-2010} out of the FMO complex.
BChls 1 and 6 are oriented towards the baseplate protein as indicated by 
experimental results \cite{opticalpropertiesFMO-review-2010,FMO-orientation-PNAS-2009}. 
These sites are therefore considered as the entrance sites which are
the initially excited sites.

The surrounding vibrational pigment-protein-solvent environment induces 
thermal fluctuations on the excitation transfer dynamics. We treat
the electronic states of the FMO complex within an open quantum system 
approach \cite{Weiss-2008}. The thermal fluctuations are generated  by
environmental harmonic modes \cite{May-Kuehn-2011}. Thus the total Hamiltonian is
\begin{eqnarray}\label{eq:full-Hamiltonian}
 H = H_{FMO} &+& \sum^{7}_{j=1} |j \rangle \langle j| \sum_{\alpha} \kappa^{(j)}_{\alpha} q_{j,\alpha} \\
 &&  + \sum^{7}_{j=1} \frac{1}{2} \sum_{\alpha} \left( p^2_{j,\alpha} + \omega^2_{j,\alpha} q^2_{j,\alpha} \right), \nonumber
\end{eqnarray}
with momenta $p_{j,\alpha}$, displacements $q_{j,\alpha}$, 
frequencies $\omega_{j,\alpha}$ and coupling constants $\kappa^{(j)}_{\alpha}$ 
of the environmental vibrations at site $j$. We 
assume that the fluctuations at different BChl sites are identical in their
characteristics, but spatially uncorrelated \cite{Nalbach-PRE-2011}. Within this
open quantum system approach, the environmental fluctuations and their coupling 
to the system dynamics are characterized by their spectral density  
$G(\omega) = \sum_{j,\alpha} \left( |\kappa^{(j)}_{\alpha}|^2 / 
2\omega_{j,\alpha} \right) \delta(\omega-\omega_{j,\alpha})$. This distribution
function determines the temporal correlations of the environmental fluctuating
forces \cite{Weiss-2008}. Here, we use experimentally determined
\cite{Wendling-JPCB-2000} and theoretically parametrized
\cite{Adolphs-Renger-BJ-2006} form 
$G(\omega) = S_0 g_0(\omega) + S_H \delta(\omega-\omega_H)$.
This fluctuational spectrum contains both a broad low frequency 
contribution $S_0 g_0(\omega)$ by the protein vibrations with 
Huang-Rhys factor $S_0$ and a single effective vibrational mode 
of the pigments centered at $\omega_H = 180$ cm$^{-1}$ with 
Huang-Rhys factor $S_H=0.027$ \cite{Wendling-JPCB-2000} (the detailed form is
given in the SI). Below, we will also study the case of a stronger 
excitonic-vibrational mode coupling, since Adolphs and Renger 
have determined a larger Huang-Rhys factor of
$S_H=0.22$ \cite{Adolphs-Renger-BJ-2006}. In general, within the system-bath
approach, the environment is 
commonly assumed to be in a thermal equilibrium state. Thus, including a 
discrete vibrational mode into the environmental spectrum implicitly assumes  
that the time scale of thermalization of this mode is 
much shorter than any system time scale. Then, the discrete mode only provides
thermal equilibrium fluctuations around its thermal state. However, recent
analyses \cite{Kauffmann-JPCB-2012,Plenio-NatPhys-2013} propose that 
underdamped vibrational modes are present, which are strongly coupled to 
the electronic transitions with large Huang-Rhys factors, and which are close to resonance to excitonic transitions. The observed long-lasting coherent signals \cite{Engel-Nature-2007,Engel-PNAS-2010} are then attributed to the coupled exciton -  underdamped mode system. In this scenario, the full
nonadiabatic quantum dynamics of the vibrational mode has to be considered and not
only its thermal equilibrium fluctuations. We therefore include in the following the 
vibrational mode at $180$ cm$^{-1}$ explicitly as part of the {\it system} Hamiltonian 
and thus describe its nonequilibrium quantum dynamics on an equal footing as the excitonic 
states. Due to the exponential growth of needed computer power for QUAPI with increasing 
system size (see SI), we restrict the localized mode to its three (two) lowest energy 
eigenstates for site 3 (all others). In detail, we include all states up to the energy 
of the site with lowest energy, i.e. site 3, plus two times thermal energy.


\subsection*{Energy transfer dynamics through the FMO complex}

In order to discuss the efficiency of the energy transfer through the FMO complex towards
the RC, we model the latter as an energy sink which is 
connected solely to site $3$. We treat the transfer towards the RC as a
population decay on a purely phenomenological level by constant decay
rate of $1$ ps$^{-1}$ since we are not interested in the explicit details of 
the dumping process. Backtransfer from the RC to the FMO complex is then 
excluded here. In turn, the rise time of the population growth of the RC is
taken as a measure for the efficiency of the energy transfer through the
complex. 

\begin{figure}
\centerline{\includegraphics[width=0.45\textwidth]{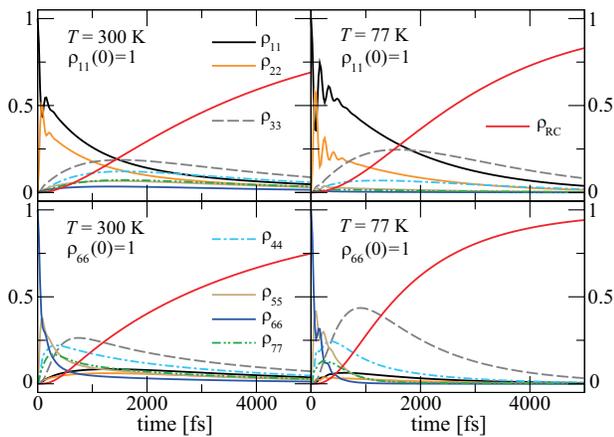}}
\caption{Time evolution of the FMO site populations and reaction center (RC) in
absence of any vibrational mode for $T=300$ K (left column) 
and $T=77$ K (right column) for the initial population of the two entrance
sites $\rho_{11}(0)=1$ (upper row) and 
$\rho_{66}(0)=1$ (lower row) for the Huang-Rhys factor
$S_H=0.027$.}\label{fig:1}
\end{figure}

To evaluate the influence of localized vibrational modes on the exciton transfer 
dynamics, we first provide the excitonic dynamics in absence of any vibrational 
mode as a reference. Fig.\ \ref{fig:1} shows the results for the individual site 
populations along with the population $\rho_{\rm RC}$ of the energy sink. We study 
two cases of the initial population of the two entrance sites, i.e.,  
$\rho_{11}(0)=1$ (upper row) and $\rho_{66}(0)=1$ (lower row) and for 
the two temperatures $T=300$ K (left column) and $T=77$ K (right column). 
We observe that apart from few oscillations of selected populations at 
very short times, no long-lasting coherent oscillations in the populations are
found. The population of the energy sink increases in a monotonous manner. 

\begin{figure}
\centerline{\includegraphics[width=0.45\textwidth]{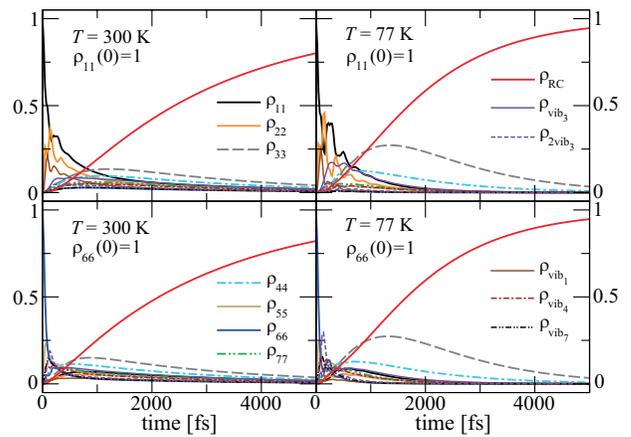}}
\caption{Same as in Fig.\ \ref{fig:1} but in presence of individual vibrational
modes at each individual molecular site. Vibrational states with energies up to
$450$ cm$^{-1}$ above the energy of site $3$ are included and all
vibrational modes have equal characteristics.}\label{fig:2}
\end{figure}

Taking the results in Fig.\ \ref{fig:1} as a reference, we next consider the 
exciton dynamics when every individual FMO site is coupled to its own vibrational 
mode. All of them are assumed to have equal characteristics and we choose in this 
section a Huang-Rhys factor of $S_H=0.027$. The QUAPI method is limited by the 
exponential growth of the array sizes and the computational times for growing system 
Hilbert space dimensions (see SI). We thus have to limit the Hilbert space 
dimensions to vibrational states with energies up to $450$ cm$^{-1}$ above 
the energy of site $3$. The exciton eigenstates of the FMO complex, obtained by 
diagonalizing the FMO system Hamiltonian, show that the included vibrational 
states have comparable or larger energies. The relevant 
vibrational modes which influence the system dynamics are those with 
energies comparable to the energy difference between the exciton
states \cite{Nalbach-JPB-2012}. Thus, the relevant FMO excitonic energy ranges
are sufficiently covered, and therefore, the technical restriction of truncated
vibrational energies has no severe implications. The results for initially excited 
entrance sites without vibrational excitation are shown in Fig.\ \ref{fig:2}. We 
observe both prolonged oscillatory population dynamics with increased coherence 
times \textit{and} an increased transfer efficiency, as indicated by the faster 
rise of the population $\rho_{\rm RC}$ of the energy sink. 

The increased transfer efficiency can be quantified in terms of the time which is 
required for the transfer of excitation energy through the FMO complex. A measure 
for this is the rise time of the exponential growth of the population $\rho_{\rm RC}$ 
of the RC to which we refer as 'transfer time' in short. The dynamical simulations
allow us to read off the rise times, the results are summarized in Table\ \ref{tab:table}. 
We find that, when vibrational modes are coupled to all individual sites separately
and including all states with energies up to $450$ cm$^{-1}$ above 
the energy of site $3$, the transfer times decreases by about $25\%$. The speed-up is slightly 
larger when the initial excitation starts at site $1$ due to the shorter 
route \cite{Brixner-Nature-2005}. In the site representation, this proceeds along 
the chain $1 \rightarrow 2 \rightarrow 3$. In turn, when the excitation starts at 
site $6$, it follows $6 \rightarrow (5/7) \rightarrow 4 \rightarrow 3$. A speed-up 
of about $25\%$ is significant.

\begin{table}
\caption{Excitation energy transfer times at $T=300$ K without and with  
nonequilibrium vibrational modes coupled to the excitonic transitions. A
negative change in the transfer time indicates a speed-up, while a positive
sign indicates a slower transfer as compared to the case
without vibrational states.}\label{tab:table}
\begin{tabular}{cccc}
 Localized vibrational & Initial excitation & Transfer time &    Change   \\
    mode coupled to    &      at site       &      [ps]     &      by     \\
 \hline \hline
     NO vibration      &         $1$        &      3.84     &      --     \\ 
                       &         $6$        &      3.39     &      --     \\ 
 \hline
       all sites       &         $1$        &      2.86     &  -25.4 $\%$ \\ 
(up to $450$ cm$^{-1}$)&         $6$        &      2.60     &  -23.2 $\%$ \\ 
 \hline
       site $1$        &         $1$        &      4.12     &  + 7.4 $\%$ \\ 
                       &         $6$        &      3.62     &  + 6.9 $\%$ \\ 
 \hline
       site $3$        &         $1$        &      2.62     &  -31.6 $\%$ \\ 
                       &         $6$        &      2.49     &  -26.6 $\%$ \\ 
 \hline
       site $6$        &         $1$        &      3.95     &  + 2.9 $\%$ \\ 
                       &         $6$        &      3.56     &  + 5.0 $\%$ \\ 
\end{tabular}
\end{table}

\begin{figure}
\centerline{\includegraphics[width=0.45\textwidth]{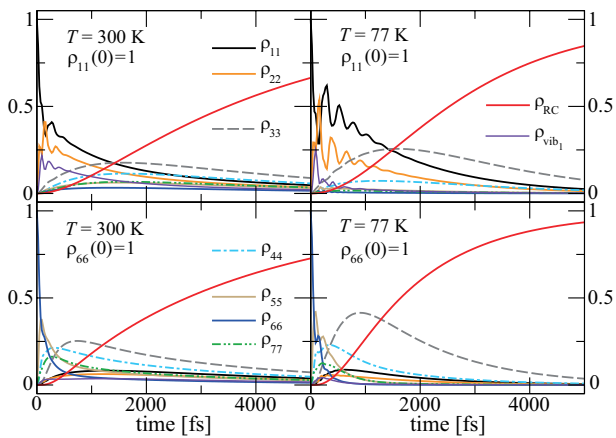}}
\caption{Time evolution of the populations of the FMO sites, the sink, and
the single vibrational mode coupled solely to the entrance site
$1$.}\label{fig:3}
\end{figure}

\begin{figure}
\centerline{\includegraphics[width=0.45\textwidth]{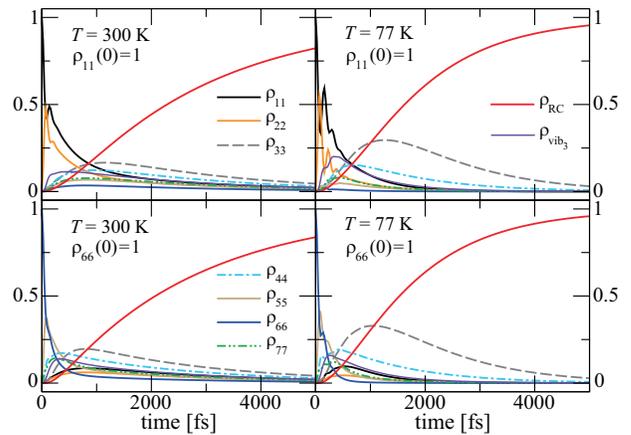}}
\caption{Same as in Fig.\ \ref{fig:3}, but with the single vibrational mode coupled
to the exit site $3$.}\label{fig:4}
\end{figure}

To further elucidate by which more detailed mechanism the vibrations enhance
the coherence times and the transfer efficiency, we study next the excitonic
dynamics by including a single localized vibrational mode only at selected
sites separately. In particular, we consider the two cases, (i) when the
vibrational mode is coupled to the entrance site $1$, and (ii) when
it is coupled to the exit site $3$. The corresponding results are presented in
Figs.\ \ref{fig:3} and \ref{fig:4}, respectively. The third case when the
vibrational mode is coupled to the other entrance site $6$ is discussed
in the SI. The resulting energy transfer times at $T=300$ K for 
all cases are given in Table\ \ref{tab:table}.

When coupling a single vibrational mode to the entrance site $1$, we observe 
in Fig.\ \ref{fig:3} for both values of the temperature, an enlarged time 
window exists with an oscillatory dynamics of the populations, after
which they continue by an incoherent decay. This effect is more pronounced when
the site $1$ is initially excited (upper row) as compared to an initial
preparation in site $6$ (lower row). In particular, the energy coherently
oscillates between the sites $1$ and $2$ over several hundreds of femtoseconds.
A closer inspection of the dynamics shows that the oscillations in 
the exciton populations indeed go back to coherent transitions between the
vibrational ground state and the vibrational first excited state at site $1$.
We extract coherence times of about $1000$ fs for $T= 77$ K and of $300$ fs
for $T= 300$ K, which coincide which those reported in the experiments
\cite{Engel-Nature-2007,Engel-PNAS-2010}. However, we observe that the
associated energy transfer times have {\em increased\/} in comparison with the
case when vibrational modes are excluded (see Table\ \ref{tab:table}). These
results prove that the coupling of a nonequilibrium vibrational mode to the
entrance site $1$ enhances coherence times, but decreases the overall transfer
efficiency. The corresponding findings also arise when a single vibrational
mode is coupled to the entrance site $6$ (see Table\ \ref{tab:table} and SI).

On the other hand, when the vibrational mode is coupled to the {\em exit\/} site
$3$, see Fig.\ \ref{fig:4}, the coherence times are not enhanced as compared to
the case without any vibrational mode (Fig.\ \ref{fig:1}). This is not
surprising since site $3$ is only rather weakly coupled to its neighbours and additionally coupled to the RC acting as an incoherent energy sink.
However, it is important to realize that the population of the sink
grows faster in this constellation and, consequently, the transfer efficiency is
increased, as seen in Table\ \ref{tab:table}. This speed-up of the transfer
efficiency is a key observation and can be directly 
rationalized in terms of an additional transfer channel which is provided by
the excited vibrational state at site $3$. This excited vibrational state is
nearly in resonance with the electronic transition and thus decreases the energy
gap with the entrance sites.
By this, it adds an additional efficient pathway in form of a vibrational decay channel into the RC. Accordingly, more states connected to the RC are available to become populated during the exciton transfer in the complex and, consequently, more states can dump their energy into the sink in parallel. The population of the sink then can grow faster and yields to an overall increased transfer efficiency. 

Overall, the underdamped modes at $180$ cm$^{-1}$ improve the efficiency of the 
quantum coherent excitation energy transfer and at the same time result in the observed 
prolongued quantum coherent oscillations. Both are a result of the modes being underdamped, 
such that they cannot thermalize on time scales fast compared to the electronic energy 
transfer dynamics. Beyond that, the speed-up of the energy transfer is rather insensitive to the actual coherence life times.


\subsection*{Population dynamics in the FMO complex}

Having studied the effect of nonequilibrium vibrational states on the exciton transfer 
efficiency, we study next the impact of a larger coupling of
the vibrational mode and the electronic transitions, i.e., for a larger 
Huang-Rhys factor. So far, we have considered an intermediate
exciton-vibrational coupling. Based on the data of the  temperature
dependence  of the fluorescence line-narrowing spectra, Wendling et al.\  
\cite{Wendling-JPCB-2000} have determined a Huang-Rhys factor $S_H
= 0.027$. This corresponds to a coupling strength of $30$ cm$^{-1}$ (see the SI
for details). On the other hand, Adolphs and Renger
\cite{Adolphs-Renger-BJ-2006} have found a Huang-Rhys factor of $S_H =
0.22$ for the vibrational mode at $180$ cm$^{-1}$, which is about one order of
magnitude larger. This corresponds to a vibrational coupling strength of
 $84$ cm$^{-1}$. These two values correspond
to the regime of the intermediate and the strong exciton-vibrational
mode coupling, respectively. We show next that for the strong coupling regime, 
the same physical picture applies as in the intermediate regime considered so
far, but small quantitative differences arise (see Table S1 in the SI).

\begin{figure}
\centerline{\includegraphics[width=0.44\textwidth]{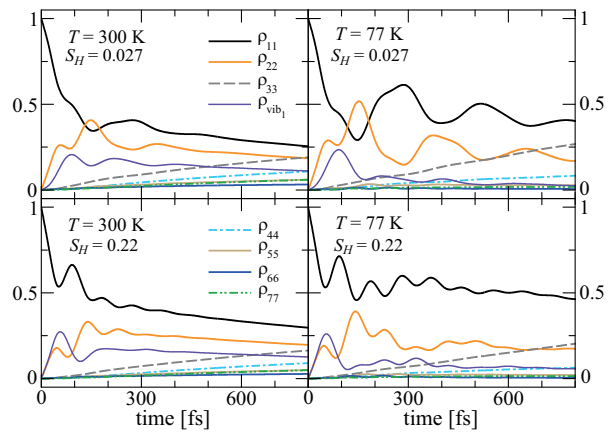}}
\caption{Time evolution of populations of the FMO sites and the first excited
state of the vibrational mode coupled to the entrance site $1$ in absence of the
reaction center. We show the results for the initial condition $\rho_{11}(0)=1$
for $T=300$ K (left column) and $T=77$ K (right column) for an intermediate
($S_H=0.027$, upper row) and a strong ($S_H=0.22$, lower row)
exciton-vibrational mode coupling.}\label{fig:5}
\end{figure}

\begin{figure}
\centerline{\includegraphics[width=0.44\textwidth]{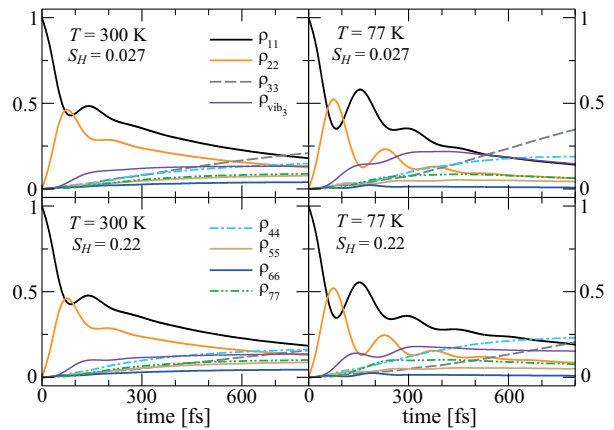}}
\caption{Same as in Fig.\ \ref{fig:5}, but with the vibrational mode coupled
to the exit site $3$.}\label{fig:6}
\end{figure}

Here, we focus on the study of the transient coherent oscillations and do
not include the additional sink at the exit site $3$. The results in Figs.\
\ref{fig:5} and \ref{fig:6} show the cases when the vibrational 
mode is coupled to the entrance site $1$ and to the exit site $3$,
respectively. The case when the vibrational mode is coupled to the
second entrance site $6$ is discussed in the SI. A single initial condition
$\rho_{11}(0)=1$ is chosen for both cases at physiological ($T=300$K) and 
cryogenic ($T=77$K) temperatures.

When the underdamped vibrational mode is strongly coupled to the initially
excited site $1$, we observe in Fig.\ \ref{fig:5} (lower row) on the one hand
strong beatings in the populations  between the vibrational ground state and the
vibrational first excited state at site $1$. On the other hand, however, we also
find coherent beatings between exciton sites $1$ and $2$ without excitation of
the vibrational state. At room temperature, coherent oscillations survive for up
to about $400$ fs and at $77$ K for at least up to $800$ fs. These 
coherence times are comparable to those of the experiments  
\cite{Engel-Nature-2007,Engel-PNAS-2010}. For an intermediate value of the
exciton-vibrational coupling, we observe in Fig.\ \ref{fig:5} (upper row) longer
intersite coherence times of around $1000$ fs. In this case, the excitonic
coupling between sites $1$ and $2$ is relatively larger than the
vibrational coupling of the vibrational ground and first excited states at site
$1$. On the other hand, in the strong coupling regime longer intrasite coherence
times are obtained because the coupling 
of the vibrational ground at site $1$ is of the same order of magnitude as the
coupling to the site $2$ or to its vibrational excited state. This competition
is reflected by the faster rise of the population of the vibrational
first excited state at site $1$.

When coupling the vibrational mode to the exit site $3$ but initially exciting
site $1$, we see in Fig.\ \ref{fig:6} that the coherence times are comparable
with those observed without explicitly taking an underdamped $180$ cm$^{-1}$ 
mode into account, see Ref. \cite{Nalbach-PRE-2011}. This is a result of
the fact that the two sites $1$ and $3$ are only very weakly coupled. At room
temperature, coherence survives for 
up to about $250$ fs and at $77$ K for up to $600$ fs at most in both regimes
of the coupling. Nevertheless, we observe a fast increase in the population of
the vibrational first excited site $3$ $\rho_{\rm vib_3}(t)$ which is  
even faster than the population accumuluation in its vibrational ground 
state $\rho_{33}(t)$. This fast population accumuluation is a result of the
refined matching of the energies of site $1$ and of  
site $3$ when exciting the vibrational state while tunneling. This is
another remarkable finding, since site $3$ is believed 
to be the exit site towards the reaction center and a larger population of site
$3$ will result in an overall increased transfer efficiency through 
the FMO complex towards the reaction center, irrespective of the excitation status of
the vibration. Thus, underdamped vibrational modes indeed improve the overall 
energy transfer efficiency.

These results show that an additional underdamped vibrational mode at site $3$
increases the transfer efficiency since it provides additional channels for the
parallel decay of the energy into the RC. In contrast, additional 
underdamped vibrational states at other sites (see SI) tend to decrease the transfer 
efficiency since they provide additional states in which the energy is
intermittently stored and eventually dissipated via the vibrational channel or
fed back into the excitonic network at a later stage. Hence, only an
efficiently connected exit site helps to improve the global transfer, while
additional states at the intermediate sites only lead to an inefficient 
spreading-out of the energy into too many channels. 


\section*{Discussion}

We have obtained numerically exact results for the real-time
nonequilibrium quantum dynamics of the excitation energy transfer through the  
FMO complex in presence of realistic environmental fluctuations and an underdamped localized vibrational mode at $180$ cm$^{-1}$. Our
analysis shows that the coupling of the excitonic transitions to a
nearly resonant vibrational mode causes strong vibrational quantum coherent
beatings in the intrasite populations of individual pigments. At the same
time, however, the nearly resonant coupling also causes strong coherent
excitonic beatings in the intersite population transfer between different
pigments. However, prolonged coherent intersite beatings do not
necessarily lead to an enhanced transfer efficiency. We show that only the
additional vibrational states at the exit site $3$ of the FMO complex help to
speed-up the global energy transfer by providing additional transfer channels
to the exit site towards the reaction center. In fact, they lead to a speed-up
of up to $30 \%$ in the transfer times through the complex. Long-lasting
coherence in the complex results from the coupling of the vibrational modes to
particular entrance and exit sites. It is, however, not functionally necessary for 
the enormous speed-up of energy transfer which thus is a rather robust mechanism.\\
These results offer a benchmark principle which may be implemented in artificial
light-harvesting systems as well. Their global quantum transfer efficiency can
be significantly increased by engineering the corresponding almost resonant 
vibrational modes, thereby maintaining the number of absorbing photoactive
sites constant.


\begin{acknowledgments}
We acknowledge financial support by the DFG Sonderforschungsbereich 925 ``Light-induced dynamics 
and control of correlated quantum systems'', by the German Academic Exchange Service (DAAD) and 
by the excellence cluster 'The Hamburg Centre for Ultrafast Imaging - Structure, Dynamics and 
Control of Matter at the Atomic Scale' of the Deutsche Forschungsgemeinschaft. 
\end{acknowledgments}




\end{document}